# Non-Abelian evolution of electromagnetic waves in a weakly anisotropic inhomogeneous medium

K.Yu. Bliokh[1,2*], D.Yu. Frolov[1], and Yu.A. Kravtsov[3,4]

[1] *Institute of Radio Astronomy, 4 Krasnoznamyonnaya St., Kharkov 61002, Ukraine*
[2] *A. Usikov Institute of Radiophysics and Electronics, 12 Akademika Proskury St., Kharkov 61085, Ukraine*
[3] *Space Research Institute, Profsoyuznaya St. 82/34, Moscow 117997, Russia*
[4] *Institute of Physics, Maritime University of Szczecin, 1-2 Waly Chrobrego St., Szczecin 70500, Poland*

A theory of electromagnetic wave propagation in a weakly anisotropic smoothly inhomogeneous medium is developed, based on the quantum-mechanical diagonalization procedure applied to Maxwell equations. The equations of motion for the translational (ray) and intrinsic (polarization) degrees of freedom are derived *ab initio*. The ray equations take into account the optical Magnus effect (spin Hall effect of photons) as well as trajectory variations owing to the medium anisotropy. Polarization evolution is described by the precession equation for the Stokes vector. In generic case, the evolution of wave turns out to be non-Abelian: it is accompanied by mutual conversion of the normal modes and periodic oscillations of the ray trajectories analogous to electron *zitterbewegung*. The general theory is applied to examples of wave evolution in media with circular and linear birefringence.

## I. INTRODUCTION

Evolution of electromagnetic waves in weakly anisotropic inhomogeneous media is of significant theoretical and practical interest for numerous problems of physics: light propagation in deformed fibers and liquid crystals, electromagnetic waves in interstellar gravitational field, microwaves in a weakly magnetized plasma, wave phenomena in condensed matter physics, etc. Similar problems are also characteristic for acoustic wave propagation in weakly anisotropic elastic media. Mathematically, the problem of wave propagation in a smoothly inhomogeneous and weakly anisotropic medium implies perturbations in two small parameters. The first, anisotropy parameter is

$$\mu_A = \frac{\|\hat{\Delta}\|}{\varepsilon_0} \ll 1, \qquad (1)$$

where $\hat{\Delta}$ is the small anisotropic part of the dielectric tensor with $\varepsilon_0$ being its isotropic part. The second one is the geometrical optics small parameter

$$\mu_{GO} = \frac{\lambda}{L} \ll 1, \qquad (2)$$

where $\lambda$ is the wavelength, whereas $L$ is the characteristic scale of the medium inhomogeneity. Equation (2) enables one to make use of the geometrical optics approach.

---

[*] E-mail: k_bliokh@mail.ru

The geometrical optics of isotropic ($\mu_A = 0$) smoothly inhomogeneous mediums is characterized by the polarization degeneracy: in the zero approximation in $\mu_{GO}$, the transverse waves of different polarizations obey the same dispersion relation and propagate along the same trajectories (rays), interfering with each other [1]. Any two orthogonal polarizations can be chosen as eigenmodes in this approximation. On the contrary, in essentially anisotropic media ($\mu_A \sim 1$), there are uniquely defined independent eigenmodes with orthogonal polarizations which propagate along different rays and do not interact with each other. The intermediate region of weak anisotropy, $\mu_A \ll 1$, is nontrivial. Even negligibly small anisotropy formally lifts the polarization degeneracy and specifies eigenmodes in the problem. At the same time, the weakness of anisotropy allows one to consider two eigenwaves as essentially interfering with each other as in the isotropic case, since the distance between mathematical rays is usually much less than the actual width of the wave beam. The closeness of dispersing characteristics of eigenmodes and the medium inhomogeneity ensure effective resonant interaction and mutual transformation of eigenmodes. Thus, in the weakly anisotropic inhomogeneous medium, the eigenmodes become coupled with each other.

An effective method describing waves in a weakly anisotropic inhomogeneous medium – the quasi-isotropic approximation of geometrical optics – has been developed in [2]. The basic achievement of this method is coupled equations for polarization evolution along the ray which take into account the influence of the medium anisotropy as well as the Rytov law of the polarization plane rotation in isotropic inhomogeneous medium. The latter effect appears in the first approximation in $\mu_{GO}$ and provides the parallel transport of the electric filed vector along the ray, which is related to the Berry phase [3]. In fact, the quasi-isotropic approximation uses the first-order approximation in parameters (1) and (2) in the equation for polarization evolution, but only the zero-order approximation in the ray equations [4]. At the same time, recent studies have shown, that even in isotropic smoothly inhomogeneous medium the ray equations acquire additional terms of the first order in $\mu_{GO}$ [5–11]. These terms are caused by the spin-orbit interaction of photons (also responsible for the Berry phase) [5,8] and provide for the conservation of the total angular momentum of a wave, including its spin part [8,9]. Due to spin-orbit interaction of photons a smoothly inhomogeneous isotropic medium can be considered as a weakly anisotropic one in the first approximation in $\mu_{GO}$ [6]. The gradient of inhomogeneity specifies a particular direction in the medium and causes weak circular birefringence. This effect is known as the optical Magnus effect, or, alternatively, as the spin Hall effect of photons or the topological spin transport of photons [5–11].

In the present paper we suggest a general theory for electromagnetic wave propagation in a weakly anisotropic inhomogeneous medium, based on the quantum mechanical diagonalization procedure applied to Maxwell equations and on the Berry phase theory. The approach consistently accounts terms of the first order in $\mu_A$ and $\mu_{GO}$ both in the equation for polarization evolution and in the ray equations, thereby generalizing the quasi-isotropic approximation. The derived equations describe evolution of center of wave packet or beam in the medium. The distinctive feature of weakly anisotropic media as compared with the isotropic case is a non-Abelian evolution of the wave polarization. It results in the lack of basis of fixed eigenmodes, non-integrability of the polarization evolution equation, and, as a consequence, a mutual transformation of the normal modes. Unlike the isotropic case, the evolution of photons in a weakly anisotropic medium seems to resemble the evolution of massive particles with a spin, e.g., electrons [8,12]. As we will show, the equation for polarization evolution reminds the Bargmann–Michel–Telegdi equation describing the Thomas precession of pseudospin (Stokes vector) of the wave. The derived ray equations involve corrections terms giving rise to deflections of rays due to both the proper medium anisotropy and the optical Magnus effect. Because of periodical changes of the wave polarization due to the mutual conversion of modes, the ray trajectories can experience oscillatory variations, similar to *zitterbewegung* of electron with spin-orbit interaction. The general theory is illustrated by characteristic examples of the wave evolution in media with circular and linear birefringence.

It is worth noticing that an alternative but related approach, considering the electromagnetic wave evolution in a gravitational field within the Bargmann–Wigner equations, has been offered recently in [13]. Our approach is in essence equivalent to that developed in [14] for electron wave-packet evolution in coupled bands. In our case non-Abelian evolution appears due to anisotropic correction in the Hamiltonian in the presence of Abelian Berry gauge field.

## II. GENERAL THEORY

### A. Diagonalization of Maxwell equations

We will consider evolution of a monochromatic electromagnetic wave packet or beam in an inhomogeneous anisotropic lossless medium. Maxwell equations for the wave electric field **E** read

$$\left[ \lambdabar^2 \, \text{curl}\,\text{curl} - \hat{\varepsilon} \right] \mathbf{E} = 0 , \tag{3}$$

where $\hat{\varepsilon}$ is the Hermitian dielectric tensor (we will mark all matrix values with hats), $\lambdabar \equiv \lambda / 2\pi = c/\omega$ is the wavelength in vacuum divided by $2\pi$, and $\omega$ is the frequency. Analogously to [10], we introduce the dimensionless differential operator of momentum

$$\mathbf{p} = -i\lambdabar \frac{\partial}{\partial \mathbf{R}} \tag{4}$$

($\mathbf{R}$ are the coordinates), which leads to commutation relations similar to the quantum mechanical ones:

$$[R_i, p_j] = i\lambdabar \delta_{ij} . \tag{5}$$

Then, equation (3) takes the form

$$-\mathbf{p} \times (\mathbf{p} \times \mathbf{E}) - \hat{\varepsilon} \mathbf{E} = 0 , \tag{6}$$

or,

$$\hat{H} \mathbf{E} = 0 , \quad \text{where} \quad \hat{H} = p^2 - \hat{Q} - \hat{\varepsilon} \tag{7}$$

is the effective Hamilton operator and $Q_{ij} = p_i p_j$. In Eq. (7) and in what follows, scalars (when they are summed up with matrices) are assumed to be multiplied by the unit matrix of the corresponding rank.

In weakly anisotropic medium the dielectric tensor $\hat{\varepsilon}$ can be represented as a sum of the main, scalar isotropic component $\varepsilon_0$ and the small matrix correction $\hat{\Delta}$ related to the medium anisotropy:

$$\hat{\varepsilon} = \varepsilon_0 + \hat{\Delta} , \tag{8}$$

where $\varepsilon_0 = \varepsilon_0(\mathbf{R})$ and $\hat{\Delta} = \hat{\Delta}(\mathbf{p}, \mathbf{R})$ [15]. Considering the wave evolution in smoothly inhomogeneous media in frame of the geometrical optics method, we will assume the first-order approximation in small parameters $\mu_A$ and $\mu_{GO}$, Eqs. (1) and (2), and neglect the second-order terms like $\mu_A \mu_{GO}$.

Equation (7) describes the eigenmodes which are mixed because of the Hamiltonian non-diagonality caused predominantly by $\hat{Q}$ matrix. It is possible to diagonalize it by a local rotation transformation superposing $z$ axis with the direction of **p** vector [10]:

$$\mathbf{E} = \hat{U}(\mathbf{p}) \mathbf{E}' , \quad \hat{U} = \begin{pmatrix} \sin\phi & \cos\theta\cos\phi & \sin\theta\cos\phi \\ -\cos\phi & \cos\theta\sin\phi & \sin\theta\sin\phi \\ 0 & -\sin\theta & \cos\theta \end{pmatrix} . \tag{9}$$

Here $(\theta, \phi)$ are spherical coordinates of the unit vector $\mathbf{p}/p$ in momentum $\mathbf{p}$-space: $\mathbf{p}/p = (\sin\theta\cos\phi, \sin\theta\sin\phi, \cos\theta)$. In the geometrical optics approach, which implies substitution of the operator $\mathbf{p}$ with 'classical' momentum $p = \hbar k$ (where $k$ is the wave vector of the center of wave packet), the transformation (9) is equivalent to the transition to the ray coordinates $\tilde{\mathbf{R}}$ locally related to the fixed coordinate frame, $\mathbf{R}$, as $\tilde{\mathbf{R}} = \hat{U}^\dagger(p)(\mathbf{R} - r)$ (where $r$ is the radius-vector of the wave packet center) [16]. Transformation (9) generates $\mu_{GO}$-order terms and, therefore, one can consider a coordinate frame attached to the zero-approximation ray, $p = p^{(0)}$, $r = r^{(0)}$ (see Subsection II D). Throughout the paper, the wave polarization is considered in this coordinate frame.

Transformation (9) yields Hamiltonian $\hat{H}' = \hat{U}^\dagger \hat{H} \hat{U}$:
$$\hat{H}' = p^2 - \hat{Q}' - \hat{U}^\dagger \varepsilon_0 \hat{U} - \hat{U}^\dagger \hat{\Delta} \hat{U}. \tag{10}$$
Here, the first three terms are the same as in the isotropic case [10], i.e. $\hat{Q}' = \hat{U}^\dagger \hat{Q} \hat{U} = \mathrm{diag}(0, 0, p^2)$, whereas the third term is transformed nontrivially because of the noncommutativity of momentum and coordinates, Eq. (5):
$$\hat{U}^\dagger(\mathbf{p}) \varepsilon_0(\mathbf{R}) \hat{U}(\mathbf{p}) = \varepsilon_0(\mathbf{R} + \hbar \hat{\mathbf{A}}) \equiv \varepsilon_0(\hat{\mathbf{R}}'). \tag{11}$$
Here
$$\hat{\mathbf{A}}(\mathbf{p}) = i \hat{U}^\dagger \frac{\partial \hat{U}}{\partial \mathbf{p}} \tag{12}$$
is a pure gauge potential in the $\mathbf{p}$-space, induced by the local gauge transformation (9) and
$$\hat{\mathbf{R}}' = \mathbf{R} + \hbar \hat{\mathbf{A}} \tag{13}$$
is the operator of covariant coordinates corresponding to the center of the semiclassical wave packet [7,8,10].

The fourth term in Eq. (10) characterizing the medium anisotropy, depends on noncommuting coordinates and momentum. However, owing to the smallness of anisotropy it is proportional to $\mu_A$, so that one can neglect commutators proportional to $\mu_{GO}$ [15] and multiply operators as usual matrices: $\hat{\Delta}' = \hat{U}^\dagger \hat{\Delta} \hat{U}$.

As a result, the wave equation and Hamiltonian become
$$\hat{H}' \mathbf{E}' = 0, \quad \hat{H}'(\mathbf{p}, \hat{\mathbf{R}}') = p^2 - \hat{Q}' - \varepsilon_0(\hat{\mathbf{R}}') - \hat{\Delta}'(\mathbf{p}, \hat{\mathbf{R}}'). \tag{14}$$
Dealing with usual canonical coordinates, in the first approximation in parameters $\mu_A$ and $\mu_{GO}$ one has
$$\hat{H}'(\mathbf{p}, \mathbf{R}) = p^2 - \hat{Q}' - \varepsilon_0(\mathbf{R}) - \hbar \hat{\mathbf{A}}(\mathbf{p}) \nabla \varepsilon_0(\mathbf{R}) - \hat{\Delta}'(\mathbf{p}, \mathbf{R}). \tag{15}$$
Here and in what follows we take into account that the difference between usual and covariant coordinates can be neglected in the first-order terms [15]. Hamiltonian (15) is almost block-diagonal; its upper left $2 \times 2$ sector describes almost-transverse electromagnetic waves, whereas the lower right element with index 33 corresponds to the longitudinal wave. The latter can exist near resonance only, when $\varepsilon_0 = 0$, and will be excluded from further analysis. As follows from the adiabatic theory, the small cross elements with indices 13, 23, 31, and 32, contained in the last two terms of Eq. (15), can be neglected as they make the second-order contribution to the evolution of waves. Therefore, when dealing with the transverse waves, one can consider only upper left $2 \times 2$ sector of equations (14) and (15). Denoting this sector of operators $\hat{H}'$, $\hat{\Delta}'$, $\hat{\mathbf{R}}'$, $\hat{\mathbf{A}}$ and two upper, transverse components of the field $\mathbf{E}'$ as $\hat{h}$, $\hat{\delta}$, $\hat{\mathbf{r}}$, $\mathcal{\hat{A}}$, and $\mathbf{e}$, respectively, we arrive at
$$\hat{h}\mathbf{e} = 0, \quad \hat{h} = \frac{1}{2}\left[p^2 - \varepsilon_0(\hat{\mathbf{r}}) - \hat{\delta}\right] = \frac{1}{2}\left[p^2 - \varepsilon_0(\mathbf{R}) - \hbar \mathcal{\hat{A}} \nabla \varepsilon_0 - \hat{\delta}\right], \tag{16}$$

where the factor 1/2 is introduced for the convenience of what follows. A transition from the equations (14) and (15) to the reduced equation (16) is equivalent to the projection of the 3-dimensional electric field vector on the plane perpendicular to $\mathbf{p}$, i.e. to the ray.

Hereinafter we will use the basis of circularly polarized waves, which diagonalizes the spin-orbit interaction of photons [6–10]. By not introducing new notations, we realize transition to this basis by a global transformation

$$\mathbf{e} \to \hat{V}\mathbf{e},\ \hat{h} \to \hat{V}^\dagger \hat{h} \hat{V},\ \hat{\mathcal{A}} \to \hat{V}^\dagger \hat{\mathcal{A}} \hat{V},\ \hat{\delta} \to \hat{V}^\dagger \hat{\delta} \hat{V}, \qquad (17)$$

where $\hat{V} = \dfrac{1}{\sqrt{2}}\begin{pmatrix} 1 & 1 \\ i & -i \end{pmatrix}$.

### B. Berry connection and curvature

The potential $\hat{\mathcal{A}}$ is not a pure gauge one anymore and a nonzero field $\hat{\mathcal{F}} = \dfrac{\partial}{\partial \mathbf{p}} \times \hat{\mathcal{A}}$ corresponds to it. These are the Berry gauge potential and field (or Berry connection and curvature), which describe the parallel transport of the electric field vector along the ray, Berry phase, and topological spin transport of photons [3,6–10,17]. The presence of non-zero Berry curvature is directly related to the noncommutativity of the operators of covariant coordinates for transverse waves [8,10,17]:

$$[\hat{r}_i, \hat{r}_j] = i \lambdabar^2 e_{ijk} \hat{\mathcal{F}}_k, \qquad (18)$$

where $e_{ijk}$ is the unit antisymmetric tensor. Direct calculations from Eqs. (9), (12), and (17) yield [10]

$$\hat{\mathcal{A}} = \mathbf{a}\hat{\sigma}_z,\ \hat{\mathcal{F}} = \mathbf{f}\hat{\sigma}_z, \qquad (19)$$

where

$$\mathbf{a} = p^{-1} \cot\theta \left( -\sin\phi, \cos\phi, 0 \right),\ \mathbf{f} = -\dfrac{\mathbf{p}}{p^3}, \qquad (20)$$

and $\hat{\sigma}_z = \mathrm{diag}(1,-1)$ is the Pauli matrix. The Berry connection and curvature are proportional to single Pauli matrix $\hat{\sigma}_z$, which evidences Abelian nature of these fields and the evolution determined by them.

In virtue of Eqs. (19) and (20), the evolution of transverse electromagnetic waves occurs in the effective field of the 'magnetic monopole' located in the origin of momentum space, which takes opposite signs for right-hand and left-hand circularly-polarized waves [6–10]. The diagonality of the Berry connection and curvature in the basis of circularly-polarized waves ensures independence of these modes in an isotropic smoothly inhomogeneous medium. On the contrary, in a weakly anisotropic medium, the correction $\hat{\delta}$ in Eq. (16) is non-diagonal in general case, which leads to coupling and transformations of two circular modes and evidences non-Abelian nature of polarization evolution in weakly anisotropic inhomogeneous media.

### C. Pseudospin

The last two terms in the right-hand side of Eq. (16) are Hermite matrix operators which determine the polarization evolution of the wave. They can be expanded on the basis of Pauli matrices $\hat{\boldsymbol{\sigma}}$:

$$-\dfrac{1}{2}\left[ \lambdabar \hat{\mathcal{A}} \nabla \varepsilon_0 + \hat{\delta} \right] = \lambdabar \boldsymbol{\alpha} \hat{\boldsymbol{\sigma}}. \qquad (21)$$

[Generally speaking, the tensor $\hat{\delta}$ can also contain a scalar correction proportional to the unit matrix. However, such a correction can always be ascribed to the main scalar permittivity $\varepsilon_0$ (see example of Subsection III B).] Operator $\hat{\boldsymbol{\sigma}}$ can be treated as 'pseudospin' of the problem [18] with vector $\boldsymbol{\alpha} = \boldsymbol{\alpha}(\mathbf{p},\mathbf{R})$ being the 'effective magnetic field' affecting its evolution ($\lambdabar$ replaces the Planck constant in this analogy). It will be clear below that pseudospin $\hat{\boldsymbol{\sigma}}$ corresponds to the polarization Stokes vector.

The first term in Eq. (21) represents the spin-orbit interaction of photons in isotropic inhomogeneous medium: $\hat{h}_{SOI} = -\lambdabar \hat{\mathcal{A}} \nabla \varepsilon_0 / 2 \equiv \lambdabar \boldsymbol{\alpha}_{SOI} \hat{\boldsymbol{\sigma}}$. The known spin-orbit interaction term for electron can be represented in an absolutely similar form [8,12]. It follows from Eq. (19) that $\boldsymbol{\alpha}_{SOI} = (0, 0, -a \nabla \varepsilon_0 / 2)$. The second term in Eq. (21) is related to anisotropy of the medium and can be treated as the Zeeman term with 'magnetic field' $\boldsymbol{\alpha}_A$: $\hat{h}_A = -\hat{\delta}/2 = \lambdabar \boldsymbol{\alpha}_A \hat{\boldsymbol{\sigma}}$. Unlike $\boldsymbol{\alpha}_{SOI}$, vector $\boldsymbol{\alpha}_A$ generally contains not only $z$-component in the chosen basis, which determines the non-Abelian nature of the evolution related to Hamiltonian $\hat{h}_{SOI} + \hat{h}_A$.

In the introduced notations the Hamiltonian (16) takes the form

$$\hat{h} = \frac{1}{2}\left[ p^2 - \varepsilon_0(\mathbf{R}) \right] + \lambdabar \boldsymbol{\alpha} \hat{\boldsymbol{\sigma}}, \tag{22}$$

where $\boldsymbol{\alpha} = \boldsymbol{\alpha}_{SOI} + \boldsymbol{\alpha}_A$. In terms of Hamiltonian (22), the evolution is Abelian only if there is a global basis in which $\boldsymbol{\alpha} = (0, 0, \alpha)$. Otherwise polarization evolution becomes non-Abelian.

### D. Equations of motion

Quantum-mechanical approach enables one to derive the equations of motion for the wave packet or beam evolution in a straightforward way. In the Heisenberg representation, equations of motion for operators of corresponding quantities read:

$$\dot{\hat{\mathbf{p}}} = -i\lambdabar^{-1}\left[\hat{\mathbf{p}}, \hat{h}\right]_H, \quad \dot{\hat{\mathbf{r}}} = -i\lambdabar^{-1}\left[\hat{\mathbf{r}}, \hat{h}\right]_H, \quad \dot{\hat{\boldsymbol{\sigma}}} = -i\lambdabar^{-1}\left[\hat{\boldsymbol{\sigma}}, \hat{h}\right]_H. \tag{23}$$

Here the dot stands for the derivative with respect to ray parameter, 'time' $\tau$ (which will be specified at the end of this Subsection) and the subscript '$H$' indicates the Heisenberg representation for the whole equation (in particular, $\hat{\boldsymbol{\sigma}}_H$ is a 'time'-dependent operator $\hat{\boldsymbol{\sigma}}_H(\tau)$ rather than Pauli matrices). The first two equations (23) describe evolution of the translational degrees of freedom of the wave, whereas the last one corresponds to the intrinsic (spin) degree of freedom, i.e. polarization. By calculating commutators (23) with the help of Eqs. (5), (16), (18), and (22), and keeping the first-order terms in $\mu_A$ and $\mu_{GO}$, we arrive at (cf. [7–10])

$$\dot{\hat{\mathbf{p}}} = \frac{1}{2}\frac{\partial(\varepsilon_0 + \hat{\delta})}{\partial \mathbf{r}}\bigg|_H, \quad \dot{\hat{\mathbf{r}}} = \hat{\mathbf{p}} + \lambdabar \hat{\mathbf{F}} \times \dot{\hat{\mathbf{p}}} - \frac{1}{2}\frac{\partial \hat{\delta}}{\partial \mathbf{p}}\bigg|_H, \tag{24}$$

$$\dot{\hat{\boldsymbol{\sigma}}} = 2\boldsymbol{\alpha} \times \hat{\boldsymbol{\sigma}}_H, \tag{25}$$

where all the functions $\varepsilon_0$, $\hat{\delta}$, $\hat{\mathcal{F}}$, and $\boldsymbol{\alpha}$ as well as their derivatives contain $\hat{\mathbf{p}}_H$ and $\hat{\mathbf{r}}_H$ as arguments. According to the Ehrenfest theorem, equations of motion (24) and (25) also take place for the corresponding 'classical' quantities (expectation values), which can be defined as $\boldsymbol{p} = \mathbf{e}^\dagger \mathbf{p} \mathbf{e}$, $\boldsymbol{r} = \mathbf{e}^\dagger \mathbf{R} \mathbf{e}$, and $\boldsymbol{s} = \mathbf{e}^\dagger \hat{\boldsymbol{\sigma}} \mathbf{e}$, i.e. $p_i = \mathbf{e}^\dagger p_i \mathbf{e}$, etc. In so doing, $\boldsymbol{r}$ are the coordinates of the center of gravity of the wave packet, $\boldsymbol{p}$ is its momentum related to the central wave vector $\boldsymbol{k}$ as $\boldsymbol{p} = \lambdabar \boldsymbol{k}$, and $\boldsymbol{s}$ is unit vector of the classical pseudospin in the problem (as we will see, it is the Stokes vector for the central wave-packet polarization). Equations (24) and (25) for these values yield

$$\dot{\boldsymbol{p}} = \frac{1}{2}\frac{\partial(\varepsilon_0 + \delta)}{\partial \boldsymbol{r}}, \quad \dot{\boldsymbol{r}} = \boldsymbol{p} + \lambdabar \boldsymbol{\mathcal{F}} \times \dot{\boldsymbol{p}} - \frac{1}{2}\frac{\partial \delta}{\partial \boldsymbol{p}}. \tag{26}$$

$$\dot{\boldsymbol{s}} = 2\boldsymbol{\alpha} \times \boldsymbol{s}. \tag{27}$$

Here $\mathcal{F} = \mathbf{e}^{\dagger}\hat{\mathcal{F}}\mathbf{e}$, $\delta = \mathbf{e}^{\dagger}\hat{\delta}\mathbf{e}$, whereas functions $\varepsilon_0$, $\delta$, $\mathcal{F}$, and $\boldsymbol{\alpha}$ have $\boldsymbol{p}$ and $\boldsymbol{r}$ as their arguments. Since $\mathcal{F} = s_z \boldsymbol{f} = -s_z \boldsymbol{p}/p^3$ and $\delta = -2\bar{\lambda}\boldsymbol{\alpha}_A \boldsymbol{s}$, equations (26) can be rewritten as

$$\dot{\boldsymbol{p}} = \frac{1}{2}\frac{\partial \varepsilon_0}{\partial \boldsymbol{r}} - \bar{\lambda}\frac{\partial(\boldsymbol{\alpha}_A \boldsymbol{s})}{\partial \boldsymbol{r}}, \quad \dot{\boldsymbol{r}} = \boldsymbol{p} - \bar{\lambda}\left[s_z \frac{\boldsymbol{p}\times\dot{\boldsymbol{p}}}{p^3} + \frac{\partial(\boldsymbol{\alpha}_A \boldsymbol{s})}{\partial \boldsymbol{p}}\right]. \tag{26'}$$

Equations (26) and (27) are the central result of the paper. Ray equations (26) or (26') describe motion of the wave packet center in phase space $(\boldsymbol{p},\boldsymbol{r})$. In turn, equation (27) describes precession of the pseudospin in the effective field $\boldsymbol{\alpha}$ and evolution of the wave polarization. It is important to note that Eqs. (26) and (27) are essentially coupled with each other: the pseudospin evolution depends on the ray trajectory through $\boldsymbol{\alpha} = \boldsymbol{\alpha}(\boldsymbol{p},\boldsymbol{r})$ and, vice versa, the rays are perturbed by pseudospin as it can be seen from Eq. (26'). Such mutual influence of internal and externals degrees of freedom appears in the first-order approximation. The last terms in equations (26) and (26') describe polarization-dependent ray deflections due to the anisotropy, whereas the next-to-last term in the second equations (proportional to the Berry curvature) is responsible for the optical Magnus effect stemming from the spin-orbit interaction of photons [5–11]. Both the effects appear in the equations additively, which is natural in frame of the approximation linear in $\mu_{GO}$ and $\mu_A$. Note that the derived ray equations are related to the behavior of the center of gravity of the total wave packet, while it can actually be split into slightly shifted packets with different polarizations as a consequence of both circular birefringence due to the optical Magnus effect [19] and a birefringence due to the anisotropy.

Equations of motion (26) and (27) should be considered in the context of the perturbation theory in $\mu_{GO}$ and $\mu_A$. Separating zero- and first-order approximations: $\boldsymbol{r} = \boldsymbol{r}^{(0)} + \boldsymbol{r}^{(1)}$ and $\boldsymbol{p} = \boldsymbol{p}^{(0)} + \boldsymbol{p}^{(1)}$, in the zero approximation from Eqs. (26) or (26') we have:

$$\dot{\boldsymbol{p}}^{(0)} = \frac{1}{2}\nabla \varepsilon_0, \quad \dot{\boldsymbol{r}}^{(0)} = \boldsymbol{p}^{(0)}, \tag{28}$$

where $\nabla \varepsilon_0 = \nabla \varepsilon_0(\boldsymbol{r}^{(0)})$. These are traditional polarization-independent ray equations of the geometrical optics for isotropic inhomogeneous media [1]. Taking into account that Eq. (27) originates from the first-order terms, we obtain the following equations of the first approximation:

$$\dot{\boldsymbol{p}}^{(1)} = \frac{1}{2}\left[\left(\boldsymbol{r}^{(1)}\frac{\partial}{\partial \boldsymbol{r}}\right)\frac{\partial \varepsilon_0}{\partial \boldsymbol{r}} + \frac{\partial \delta}{\partial \boldsymbol{r}}\right], \quad \dot{\boldsymbol{r}}^{(1)} = \boldsymbol{p}^{(1)} + \bar{\lambda}\mathcal{F}\times\dot{\boldsymbol{p}}^{(0)} - \frac{1}{2}\frac{\partial \delta}{\partial \boldsymbol{p}}, \tag{29}$$

$$\dot{\boldsymbol{s}} = 2\boldsymbol{\alpha}\times\boldsymbol{s}. \tag{30}$$

In these equations functions $\varepsilon_0$, $\delta$, $\mathcal{F}$, $\boldsymbol{\alpha}$, as well as their derivatives, should be considered on the zero-approximation ray, i.e. at $\boldsymbol{p} = \boldsymbol{p}^{(0)}$, $\boldsymbol{r} = \boldsymbol{r}^{(0)}$. An alternative form of (29), corresponding to Eq. (26'), is

$$\dot{\boldsymbol{p}}^{(1)} = \frac{1}{2}\left(\boldsymbol{r}^{(1)}\frac{\partial}{\partial \boldsymbol{r}}\right)\frac{\partial \varepsilon_0}{\partial \boldsymbol{r}} - \bar{\lambda}\frac{\partial(\boldsymbol{\alpha}_A \boldsymbol{s})}{\partial \boldsymbol{r}}, \quad \dot{\boldsymbol{r}}^{(1)} = \boldsymbol{p}^{(1)} - \bar{\lambda}\left[s_z \frac{\boldsymbol{p}^{(0)}\times\dot{\boldsymbol{p}}^{(0)}}{p^{(0)3}} + \frac{\partial(\boldsymbol{\alpha}_A \boldsymbol{s})}{\partial \boldsymbol{p}}\right]. \tag{29'}$$

The deflections of ray trajectories, described by equations (29) or (29'), essentially depend on the pseudospin precessing according to Eq. (30). It can give rise to oscillations of the ray trajectory characterized by 'frequency' $2\boldsymbol{\alpha}$. Such oscillations are similar to those of the electron trajectory (*zitterbewegung*), which relate to interference of two close-level states split due to spin-orbit or Zeeman interaction [21]. Electron *zitterbewegung* is also directly related to its non-Abelian evolution [8,12]. In the case of electromagnetic waves (photons) *zitterbewegung* can appear only in anisotropic medium; it can be associated with the transitions between two polarization states at the conversion of modes. The effect does not arise in isotropic medium where the ray equations depend only on helicity $s_z$ (see Subsection II E) being an invariant of Eq. (30) in this case. Evolving an analogy with the electron, equation (30) is a counterpart of the Bargmann–Michel–Telegdi equation for the electron spin precession [20,12], where spin-orbit and Zeeman fields are given by $\boldsymbol{\alpha}_{SOI}$ and $\boldsymbol{\alpha}_A$, respectively.

In addition to the equations of motion, we should derive the dispersion relation, which plays the role of a constraint. Multiplying the initial wave equation (16) from the left by $2\mathbf{e}^\dagger$, one obtains the local dispersion equation for the wave packet center:

$$2\mathbf{e}^\dagger \hat{h}\mathbf{e} = p^2 - \varepsilon_0(\boldsymbol{r}) - \delta(\boldsymbol{p},\boldsymbol{r}) = 0. \tag{31}$$

In the zero approximation in $\mu_A$ it takes the form of dispersion relation for isotropic medium: $p^{(0)2} = \varepsilon_0(\boldsymbol{r}^{(0)})$. Now we can conclude that the above-introduced ray parameter $\tau$ is related to the ray length $l$ as $d\tau = dl/|\dot{\boldsymbol{r}}|$ where one has to substitute $|\dot{\boldsymbol{r}}|$ from the ray equations (26), (28), and (29), taking Eq. (31) into account. For instance, in the zero approximation $|\dot{\boldsymbol{r}}^{(0)}| = \sqrt{\varepsilon_0(\boldsymbol{r}^{(0)})}$.

### E. Polarization evolution

Let us consider a connection between pseudospin $\boldsymbol{s}$ and the wave polarization. For this purpose we turn back to the Schrödinger-type representation, Eqs. (16) and (22), and make a geometrical optics (WKB) ansatz: $\mathbf{e}(\tau) = \boldsymbol{\xi}(\tau)\exp\left(i\lambdabar^{-1}\int \boldsymbol{p}^{(0)}d\boldsymbol{r}^{(0)}\right)$. Here $\boldsymbol{\xi} = \begin{pmatrix} \xi^+ \\ \xi^- \end{pmatrix}$ is the unit complex vector of polarization of the wave packet center in the basis of circularly polarized waves, $\mathbf{e}^\dagger\mathbf{e} = \boldsymbol{\xi}^\dagger\boldsymbol{\xi} = 1$. (We do not consider here variations of the wave amplitude caused by diffraction phenomena as they do not affect the geometrical-optics characteristics of the wave.) In fact, $\boldsymbol{\xi}$ is the Johnes vector in the basis of circular polarizations. Substitution of this representation in Eqs. (16) or (22) with Eq. (28) taken into account leads, in the first-order approximation, to evolution equation for the polarization vector:

$$\dot{\boldsymbol{\xi}} = i\left[\hat{\boldsymbol{\mathcal{A}}}\dot{\boldsymbol{p}}^{(0)} + \frac{\lambdabar^{-1}}{2}\hat{\delta}\right]\boldsymbol{\xi}, \tag{32}$$

or

$$\dot{\boldsymbol{\xi}} = -i(\boldsymbol{\alpha}\hat{\boldsymbol{\sigma}})\boldsymbol{\xi}, \tag{32'}$$

where $\hat{\boldsymbol{\mathcal{A}}}$, $\hat{\delta}$ and $\boldsymbol{\alpha}$ should be taken on the zero-approximation ray, $\boldsymbol{p} = \boldsymbol{p}^{(0)}$ and $\boldsymbol{r} = \boldsymbol{r}^{(0)}$, so that $d/d\tau = \boldsymbol{p}^{(0)}\nabla$. Equation (32) or (32') describes evolution of the polarization along the ray in the Schrödinger-type representation. In isotropic medium, $\hat{\delta} \equiv 0$, Eq. (32) can be integrated owing to Abelian character of the Berry connection $\hat{\boldsymbol{\mathcal{A}}}$, Eq. (19). Equation (32) splits into two independent equations, which evidences the independence of circular modes in a smoothly-inhomogeneous isotropic medium [6–10]. The result describes the Berry phase of circularly polarized waves and Rytov law of rotation of the polarization plane [3,7,9,10,17]. In anisotropic medium, where tensor $\hat{\delta}$ is nondiagonal in general case, Eq. (32) describes non-Abelian evolution of the polarization vector $\boldsymbol{\xi}$ and the normal mode conversion.

Formally one can represent the solution of equations (32) and (32') as

$$\boldsymbol{\xi} = \mathcal{P}\exp\left[-i\int_0^\tau \boldsymbol{\alpha}\hat{\boldsymbol{\sigma}}\,d\tau'\right]\boldsymbol{\xi}_0 = \mathcal{P}\exp\left[i\int_0^\tau\left(\hat{\boldsymbol{\mathcal{A}}}\dot{\boldsymbol{p}}^{(0)} + \frac{\lambdabar^{-1}}{2}\hat{\delta}\right)d\tau'\right]\boldsymbol{\xi}_0, \tag{33}$$

where $\mathcal{P}$ is the chronological ordering operator and $\boldsymbol{\xi}_0 \equiv \boldsymbol{\xi}(0)$. The first term within the integral, which describes the Berry phase [3,7,9,10,17], can be represented as a contour integral:

$$\hat{\Theta}_B = \int_0^\tau \hat{\mathcal{A}} \dot{\boldsymbol{p}}^{(0)} d\tau' = \int_C \hat{\mathcal{A}} d\boldsymbol{p} = \hat{\sigma}_z \int_C \boldsymbol{a} d\boldsymbol{p} \equiv \hat{\sigma}_z \theta_B, \quad (34)$$

where $C = \{\boldsymbol{p}^{(0)}(\tau)\}$ is the contour of evolution of the zero-approximation ray in $\boldsymbol{p}$–space. As a result

$$\boldsymbol{\xi} = \mathcal{P} \exp\left[i\left(\hat{\Theta}_B + \frac{\lambdabar^{-1}}{2}\int_0^\tau \hat{\delta} d\tau'\right)\right] \boldsymbol{\xi}_0. \quad (35)$$

Equations (32) and (32') correspond to equation (30) for the pseudospin precession in the Heisenberg representation. Equation (30) can readily be obtained from Eq. (32') by differentiating expression $\boldsymbol{s} = \boldsymbol{e}^\dagger \hat{\boldsymbol{\sigma}} \boldsymbol{e} = \boldsymbol{\xi}^\dagger \hat{\boldsymbol{\sigma}} \boldsymbol{\xi}$. The relation between vectors $\boldsymbol{s}$ and $\boldsymbol{\xi}$ can also be obtained by means of the density matrix which equals $\hat{\rho} = \frac{1}{2}(1 + \boldsymbol{s}\hat{\boldsymbol{\sigma}})$ and, at the same time, $\rho_{ij} = \xi_i \xi_j^*$ (where $i,j = 1,2$ and $\xi_{1,2} \equiv \xi^\pm$). It follows from the above expressions that $\boldsymbol{s}$ is nothing else than the Stokes vector of pure polarized state (see [22]). Hence, Eq. (30) is an equation of the Stokes vector precession. A similar equation has been introduced earlier in [23] on the basis of simplified phenomenological assumptions, while here the equation for the Stokes vector precession is rigorously derived directly from Maxwell equations in general case. Note that $z$ component of the pseudospin, $s_z = \boldsymbol{\xi}^\dagger \hat{\sigma}_z \boldsymbol{\xi} = |\xi^+|^2 - |\xi^-|^2$, represents the mean helicity of the wave, so that the optical Magnus effect term in ray equations (26), (26'), (29) is proportional to the mean helicity. In isotropic medium, helisity is conserved in Eq. (30), $\dot{s}_z = 0$, since $\boldsymbol{s}$ precesses about $\boldsymbol{\alpha} = \boldsymbol{\alpha}_{SOI}$ which has $z$ component only.

### F. Comparison with quasi-isotropic approximation

Let us compare our equations with the quasi-isotropic approximation equations [2]. Quasi-isotropic approximation deals predominantly with polarization vector $\boldsymbol{\xi}_F$ in the linear-polarization basis attached to the Frenet normal and binormal to the zero-approximation ray, Eq. (28): $\boldsymbol{\xi}_F = \begin{pmatrix} \xi_n \\ \xi_b \end{pmatrix}$. Polarization evolution equation of the quasi-isotropic approximation reads [2]:

$$\dot{\boldsymbol{\xi}}_F = i\left[\sqrt{\varepsilon_0}\chi \hat{\sigma}_y + \frac{\lambdabar^{-1}}{2}\hat{\delta}_F\right] \boldsymbol{\xi}_F \quad (36)$$

where $\chi$ is the ray torsion and $\hat{\delta}_F$ stands for tensor $\hat{\delta}$ presented in the normal-binormal ray coordinates.

One can make sure that equations (32) and (36) are equivalent to each other. Indeed, the second terms in brackets in Eqs. (32) and (36) are coincident. Pauli matrix $\hat{\sigma}_y$ appears in Eq. (36) instead of matrix $\hat{\sigma}_z$ in Eq. (32) because of transition from the circular-polarization basis to the linear-polarization one. Finally, the integrals of the first terms in brackets in Eqs. (32) and (36) are equal to the Berry phase calculated in the respective basis. For cyclic evolutions in $\boldsymbol{p}$-space, when the contour $C$ is a loop, the following equality takes place for the Berry phase [3]:

$$\theta_B = \oint_C \boldsymbol{a} d\boldsymbol{p} = \int_S \boldsymbol{f} d^2 \boldsymbol{p} = \int_0^\tau \sqrt{\varepsilon_0} \chi d\tau' = -\Omega. \quad (37)$$

Here $S$ is the surface strained on the loop $C$ ($C = \partial S$) and $\Omega$ is the solid angle at which the surface is seen from the origin of $\boldsymbol{p}$-space. Equation (37) relates the Berry phase to the parallel transport of the electric field vector along the ray [3]. It implies that the first terms in brackets in Eqs. (32) and (36) differ by a total derivative of some function. This difference is a gauge correction because of the local rotation (about the tangent vector to a ray) of the Frenet normal-binormal ray coordinates with respect to ray coordinates used in our approach as well as in [16].

Thus, the approach presented here is completely equivalent to the quasi-isotropic approximation with regard to the polarization evolution equation. At the same time, in contrast to the quasi-isotropic approximation, which uses only zero-order ray equations (28) [4], our approach involves first-order corrections into the ray equations thereby revealing nontrivial dependence of ray trajectories on the wave polarization.

### G. Applicability conditions

Applicability of the derived equations require the terms neglected in the wave equation to be much less than the terms of the $\mu_{GO}$ and $\mu_A$ order. Inequalities $\mu_{GO}^2, \mu_A^2, \mu_{GO}\mu_A \ll \mu_{GO}, \mu_A$ lead to the following conditions for the anisotropy weakness:

$$\sqrt{\mu_{GO}} \ll \mu_A \ll \mu_{GO}^2. \tag{38}$$

Besides, the neglected terms should not lead to appreciable phase incursion and essential polarization changes, which implies restriction on the ray length $l$:

$$\frac{l}{2\pi\lambdabar}\max\left(\mu_{GO}^2, \mu_A^2\right) \ll 1. \tag{39}$$

Finally, the characteristic width of real wave packet or beam, $w$, should be large as compared with the wavelength,

$$\frac{\lambdabar}{w} \ll 1, \tag{40}$$

which enables one to make use of the paraxial (semiclassical) approximation and to associate the evolution of the beam with its central plane wave. Also, a single wave packet or beam description implies that characteristic ray deflections, $\left|r^{(1)}\right| \le \max\left(\mu_{GO}, \mu_A\right) l$, are small as compared with the beam's width:

$$\frac{\max\left(\mu_{GO}, \mu_A\right) l}{w} \ll 1. \tag{41}$$

## III. EXAMPLES

### A. Circularly-birefringent medium

In anisotropic medium with circular birefringence, the electric induction can be represented as [24]:

$$\mathbf{D} = \hat{\varepsilon}\mathbf{E} = \varepsilon_0\mathbf{E} + i\mathbf{E}\times\mathbf{g}, \tag{42}$$

where $\mathbf{g} = \mathbf{g}(\mathbf{p}, \mathbf{R})$ is the gyration vector. In this case, the anisotropic part of the dielectric tensor (8) is

$$\hat{\Delta} = ie_{ijk}g_k = i\begin{pmatrix} 0 & g_z & -g_y \\ -g_z & 0 & g_x \\ g_y & -g_x & 0 \end{pmatrix} \tag{43}$$

After carrying out transformation (9) and (10), the anisotropic part of Hamiltonian (14) takes the form

$$\hat{\Delta}' = i \begin{pmatrix} 0 & \dfrac{\mathbf{gp}}{p} & -\dfrac{[(\mathbf{g}\times\mathbf{p})\times\mathbf{p}]_z}{p^2 \sin\theta} \\ -\dfrac{\mathbf{gp}}{p} & 0 & \dfrac{(\mathbf{g}\times\mathbf{p})_z}{\sin\theta} \\ \dfrac{[(\mathbf{g}\times\mathbf{p})\times\mathbf{p}]_z}{p^2 \sin\theta} & -\dfrac{(\mathbf{g}\times\mathbf{p})_z}{\sin\theta} & 0 \end{pmatrix}. \qquad (44)$$

Then, reduction (16) and transition to the basis of circularly polarized waves, Eq. (17), yield

$$\hat{\delta} = -\frac{\mathbf{gp}}{p}\hat{\sigma}_z. \qquad (45)$$

Thus, Hamiltonian (16) and (22) and equations (24) and (32) for the wave evolution become diagonal and correspond to Abelian evolution, as in the case of isotropic medium. The effective magnetic field, Eqs. (21) and (22), is given by $\boldsymbol{\alpha} = -\dfrac{1}{2}\left(0, 0, a\nabla\varepsilon_0 - \bar{\lambda}^{-1}\mathbf{gp}/p\right)$. As a result, the Stokes vector $\mathbf{s}$ precesses around $z$ axis, Eq. (30), and the wave helicity is conserved during the evolution: $s_z = const$. The polarization evolution equations (32)–(35) can be readily integrated to give

$$\dot{\boldsymbol{\xi}} = i\left[a\dot{p}^{(0)} - \frac{\bar{\lambda}^{-1}}{2}\frac{\mathbf{gp}}{p}\right]\hat{\sigma}_3\boldsymbol{\xi}, \text{ or } \dot{\xi}^{\pm} = \pm i\left[a\dot{p}^{(0)} - \frac{\bar{\lambda}^{-1}}{2}\frac{\mathbf{gp}}{p}\right]\xi^{\pm}, \qquad (46)$$

$$\xi^{\pm} = \exp\left[\pm i\left(\theta_B + \theta_F\right)\right]\xi_0^{\pm}, \qquad (47)$$

where

$$\theta_F = -\frac{\bar{\lambda}^{-1}}{2}\int_0^{\tau}\frac{\mathbf{gp}}{p}d\tau' \qquad (48)$$

is the 'Faraday phase' acquired by the right-hand circularly polarized wave under evolution in a gyrotropic medium. For instance, in a magnetoactive medium with an external magnetic field $\mathcal{B}$, one has $\mathbf{g} = \gamma\mathcal{B}$ ($\gamma$ is a constant characterizing magnetic activity of the medium), and the phase (48) describes the Faraday effect [24]. Equation (47) shows that in a weakly anisotropic medium with circular birefringence, the polarization evolution is determined by addition of the Faraday phase to the Berry phase characteristic for isotropic medium. As a result, one deals with superposition of the Rytov [3] and Faraday effects: the polarization ellipse turns on the angle $-(\theta_B + \theta_F)$, keeping its shape unchanged.

The ray equations (26') with anisotropic part given by Eq. (45) read:

$$\dot{\mathbf{p}} = \frac{1}{2}\frac{\partial\varepsilon_0}{\partial\mathbf{r}} - \frac{s_z}{2}\frac{\partial}{\partial\mathbf{r}}\frac{\mathbf{gp}}{p}, \quad \dot{\mathbf{r}} = \mathbf{p} - \bar{\lambda}s_z\frac{\mathbf{p}\times\dot{\mathbf{p}}}{p^3} + \frac{s_z}{2}\frac{\partial}{\partial\mathbf{p}}\frac{\mathbf{gp}}{p}. \qquad (49)$$

Equations (49) contain corrections resulting both from the spin-orbit interaction of photons (optical Magnus effect or spin Hall effect of photons) [5–11] and from the Faraday-type anisotropy [24,25]. All the corrections are proportional to the wave helicity $s_z$. The ray deflections due to the two effects are summed and turn out to be competing: the anisotropy can strengthen or compensate deflections caused by the optical Magnus effect, or vice versa. In inhomogeneous magnetoactive medium with $\mathcal{B} = \mathcal{B}(\mathbf{r})$ and $\gamma = \gamma(\mathbf{r})$, equations (49) become

$$\dot{\mathbf{p}} = \frac{1}{2}\frac{\partial\varepsilon_0}{\partial\mathbf{r}} - s_z\frac{\mathbf{p}}{2p}\left(\mathcal{B}\frac{\partial\gamma}{\partial\mathbf{r}}\right), \quad \dot{\mathbf{r}} = \mathbf{p} - s_z\left[\bar{\lambda}\frac{\mathbf{p}\times\dot{\mathbf{p}}}{p^3} - \frac{\gamma}{2}\frac{\mathbf{p}\times(\mathcal{B}\times\mathbf{p})}{p^3}\right], \qquad (50)$$

where it was taken into account, that $\dfrac{\partial}{\partial\mathbf{r}}\mathcal{B} = 0$ in virtue of Maxwell equation.
11

By a way of example, let us consider ray trajectories in cylindrically symmetric waveguide medium (see [6]) with the magnetic field $\mathcal{B}$ directed along its axis $z$. In the cylindrical coordinates $(r, \varphi, z)$, one has $\varepsilon_0 = \varepsilon_0(r)$, $\gamma = \gamma(r)$, and the zero-approximation ray trajectory, Eq. (28), is a circle $p_z^{(0)} = p_r^{(0)} = 0$, $p_\varphi^{(0)} = const$, Fig. 1. Then, equations for the ray deflections, Eq. (29'), follow from Eq. (50):

$$\dot{\boldsymbol{p}}^{(1)} = 0, \quad \dot{\boldsymbol{r}}^{(1)} = -\frac{s_z}{p^{(0)}} \left( \hbar \frac{\boldsymbol{p}^{(0)} \times \dot{\boldsymbol{p}}^{(0)}}{p^{(0)2}} - \frac{\gamma \mathcal{B}}{2} \right), \quad (51)$$

where it is assumed that $\boldsymbol{p}^{(1)} = 0$. Polarization transport of rays along the magnetic field in homogeneous medium has been predicted in [24] and measured in [25], whereas the transport of rays in isotropic waveguide medium has been considered in [6] and is presented in Fig. 1a. The deflections caused by the optical Magnus effect are directed along $z$ axis, and the magnetic field can either increase or decrease transport of circularly polarized rays. In particular, when $\mathcal{B} = \frac{2\hbar}{\gamma} \frac{\boldsymbol{p}^{(0)} \times \dot{\boldsymbol{p}}^{(0)}}{p^{(0)2}} \equiv \mathcal{B}_c$, the right-hand side of the second equation (51) vanishes and rays propagate without deflections, Fig. 1b. In the case of the opposite-sign magnetic field, $\mathcal{B} = -\mathcal{B}_c$, the polarization transport along $z$ axis intensifies and deflections become twice as large as compared to the isotropic case, Fig. 1c.

Thus, magnetic field can be used as an effective tool revealing or suppressing the natural circular birefringence (the optical Magnus effect) and optical activity (the Rytov effect and Berry phase) of inhomogeneous medium, which are caused by the spin-orbit interaction of photons.

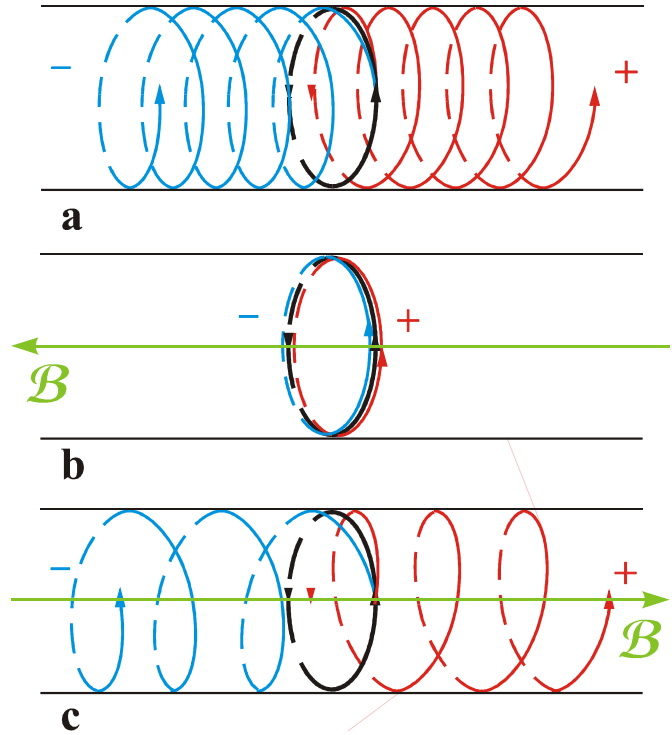

**Fig. 1.** (Color online) Ray trajectories of right-hand, "+", and left-hand, "−", circularly polarized waves in a circular waveguide at different values of an external magnetic field: $\mathcal{B} = 0$, $\mathcal{B}_c$, and $-\mathcal{B}_c$ for pictures (a), (b), and (c), respectively. The bold line depicts the zero-approximation ray.

## B. Linearly-birefringent medium

Let us consider the wave evolution in anisotropic medium with linear birefringence of the uniaxial crystal type. Such anisotropy might be induced, e.g., by an external electric field $\mathcal{E}$. The anisotropic part in the dielectric tensor (8) in this case takes the form [24]

$$\hat{\Delta} = \beta \mathcal{E}_i \mathcal{E}_j, \tag{52}$$

where $\beta$ is a scalar constant. Applying transformations (9) and (10), we find the anisotropic part of Hamiltonian (14):

$$\hat{\Delta}' = \begin{pmatrix} a^2 & ab & ac \\ ab & b^2 & bc \\ ac & bc & c^2 \end{pmatrix}, \tag{53}$$

where

$$a = \beta \frac{(\mathcal{E} \times \mathbf{p})_z}{\sqrt{p_x^2 + p_y^2}}, \quad b = \beta \frac{\left[ (\mathcal{E} \times \mathbf{p}) \times \mathbf{p} \right]_z}{p \sqrt{p_x^2 + p_y^2}}, \quad c = \beta \frac{\mathcal{E} \mathbf{p}}{p}. \tag{54}$$

After reduction (16) and transition to the basis of circularly polarized waves, Eqs. (17), we arrive at

$$\hat{\delta} = \begin{pmatrix} a^2 + b^2 & a^2 - b^2 - 2iab \\ a^2 - b^2 + 2iab & a^2 + b^2 \end{pmatrix} \tag{55}$$

Unlike the case of circular-birefringent anisotropic medium, the Hamiltonian (16), (22) with $\hat{\delta}$ from Eq. (55) is non-diagonal. By expanding $\hat{\delta}$ on the basis of Pauli matrices, we find that $\hat{\delta} = (a^2 + b^2) + (a^2 - b^2)\hat{\sigma}_x + 2ab\hat{\sigma}_y$. One can ascribe scalar correction $(a^2 + b^2)$ to the main permittivity, $\varepsilon_0 \to \varepsilon_0 + (a^2 + b^2)$, and then

$$\hat{\delta} = (a^2 - b^2)\hat{\sigma}_x + 2ab\hat{\sigma}_y. \tag{56}$$

Vector $\boldsymbol{\alpha}$, Eq. (21), which determines precession of the Stokes vector, Eq. (30), is given by

$$\boldsymbol{\alpha} = -\frac{1}{2}\left[ \bar{\lambda}^{-1}(a^2 - b^2), 2\bar{\lambda}^{-1}ab, a\nabla\varepsilon_0 \right]. \tag{57}$$

Hamiltonian (22) with Eq. (57) contains, in a generic case, all the Pauli matrices and generates non-Abelian evolution of the wave. In other words, there is no global basis of independent normal modes in the medium under consideration and mutual transformation of modes takes place.

By a way of example, let us consider the wave propagation along a helical trajectory. Such trajectories can be realized inside a cylindrical multimode waveguide (considered in the previous Subsection) as well as in coiled single-mode optical fibre (see also [26]). Let the electric field be directed along the helix axis $z$, Fig. 2. The equation of the zero-approximation ray trajectory can be set as (superscripts "(0)" are omitted):

$$x = r\cos\left[ r^{-1}\sqrt{\varepsilon_0}(\sin\vartheta)\tau \right], \quad y = r\sin\left[ r^{-1}\sqrt{\varepsilon_0}(\sin\vartheta)\tau \right], \quad z = \sqrt{\varepsilon_0}(\cos\vartheta)\tau, \tag{58}$$

where $r$ is the helix radius and $\vartheta$ is the angle between the tangent to the ray and $z$ axis. According to Eq. (28), $\mathbf{p} = \dot{\mathbf{r}}$:

$$p_x = -\sqrt{\varepsilon_0}\sin\vartheta\sin\left[ r^{-1}\sqrt{\varepsilon_0}(\sin\vartheta)\tau \right], \quad p_y = \sqrt{\varepsilon_0}\sin\vartheta\cos\left[ r^{-1}\sqrt{\varepsilon_0}(\sin\vartheta)\tau \right], \quad p_z = \sqrt{\varepsilon_0}\cos\vartheta. \tag{59}$$

As it should be, Eq. (59) satisfies the zero-approximation dispersion relation, $p^2 = \varepsilon_0$, and the first equation (28) yields $\nabla\varepsilon_0 = 2\dot{\mathbf{p}}$. By substituting Eqs. (58) and (59) into Eq. (57) with Eqs. (20) and (54) we obtain $a = 0$, $b = -\beta\mathcal{E}|\sin\vartheta|$, and

$$\boldsymbol{\alpha} = \frac{1}{2}\left(\lambdabar^{-1}\beta\mathcal{E}^2 \sin^2\vartheta, 0, -\sqrt{\varepsilon_0}\, r^{-1} \sin 2\vartheta\right). \tag{60}$$

Vector $\boldsymbol{\alpha}$ is independent of $\tau$ for the ray under consideration. Hence, according to Eq. (30), the Stokes vector uniformly precesses under the wave propagation about constant vector (60) with angular frequency

$$2\alpha = \sqrt{\left(\lambdabar^{-1}\beta\mathcal{E}^2 \sin^2\vartheta\right)^2 + \left(\sqrt{\varepsilon_0}\, r^{-1} \sin 2\vartheta\right)^2}. \tag{61}$$

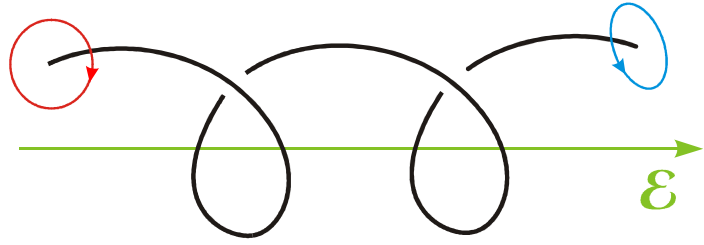

**Fig. 2.** (Color online) Helical ray of the zero approximation in an external electric field resulting in the uniaxial-crystal-type anisotropy. By measuring input and output polarizations it is possible to observe conversion of modes in the system.

It is easy to calculate variations of polarization by integrating equation (30). Since $\boldsymbol{\alpha} = (\alpha_x, 0, \alpha_z)$, let us perform the rotation transformation about $y$ axis which superposes vector $\boldsymbol{\alpha}$ with $z$ axis. If $\psi$ is the angle between $\boldsymbol{\alpha}$ and $z$ axis, so that

$$\alpha_x = \alpha \sin\psi,\ \alpha_z = \alpha \cos\psi, \tag{62}$$

the transformation is realized by the following substitution in Eq. (30):

$$\boldsymbol{s} = \hat{R}\boldsymbol{s}',\ \hat{R} = \begin{pmatrix} \cos\psi & 0 & \sin\psi \\ 0 & 1 & 0 \\ -\sin\psi & 0 & \cos\psi \end{pmatrix}. \tag{63}$$

As a result, Eq. (30) takes the form

$$\dot{\boldsymbol{s}}' = 2\lambdabar^{-1}\boldsymbol{\alpha}' \times \boldsymbol{s}', \tag{64}$$

where $\boldsymbol{\alpha}' = (0,0,\alpha)$. One can write out solution of this equation with normalization $\boldsymbol{s}'^2 = 1$ and general initial conditions $\boldsymbol{s}'(0) = (A, B, C)$ ($A^2 + B^2 + C^2 = 1$):

$$\begin{aligned} s'_x &= A\cos(2\alpha\tau) - B\sin(2\alpha\tau), \\ s'_y &= A\sin(2\alpha\tau) + B\cos(2\alpha\tau), \\ s'_z &= C. \end{aligned} \tag{65}$$

These expressions, together with Eqs. (62) and (63), completely describe the precession of the Stokes vector thereby representing the polarization evolution of the wave.

Figure 3 shows variations of the wave polarization indicated by the Stokes vector on the Poincaré sphere. Due to the Stokes vector precession, the wave polarization undergoes periodic changes. In the case of isotropic inhomogeneous medium the Stokes vector moves along parallels on the Poincaré sphere with the wave helicity conserved [3] (the circularly polarized eigenmodes), whereas in the case of uniaxial homogeneous medium the Stokes vector moves along meridians and polarization ellipse keeps its orientation unchanged (linearly polarized independent modes). In the example under discussion both eccentricity and orientation of the polarization vary. This evidences the mutual conversion of modes and energy exchange between them. Depending on the initial polarization state and direction of $\boldsymbol{\alpha}$, the polarization can change the helicity sign. Note that when the initial polarization state corresponds to the Stokes vector $\mathbf{s}(0) = \pm \boldsymbol{\alpha}/\alpha$, the polarization remains unchanged along the ray and transformation does not occur. However, such polarization states are individual for each given ray. Therefore, there is no global basis of independent eigenmodes in the system. As far as we know, the above example points out for the first time the possibility of the mode conversion due to concurrence of the circular birefringence related to the spin-orbit interaction of photons and the simplest linear birefringence of the uniaxial crystal type. Usually, mode transformation arises due to a complex anisotropy of the medium (e.g., concurrence between the Faraday circular birefringence and the Cotton−Muton linear birefringence [2]).

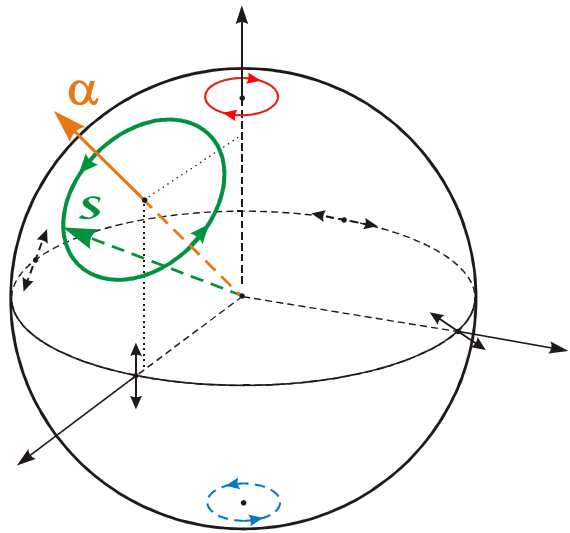

**Fig. 3.** (Color online) Precession of Stokes vector $\mathbf{s}$ on Poincaré sphere about vector $\boldsymbol{\alpha}$, Eq. (60), with "frequency" $2\alpha$, Eq. (61).

Finally, the first-approximation ray equations in the medium under consideration take the form of Eqs. (26), (26'), (29), and (29') with $\boldsymbol{\alpha}_A = -\frac{\lambdabar^{-1}}{2}\left(a^2 - b^2, 2ab, 0\right)$ and Eq. (54). They essentially depend on different components of the precessing Stokes vector $\mathbf{s}$. Hence, oscillations of the ray trajectory, *zitterbewegung*, with frequency $2\alpha$, Eq. (61), can be observed in this system.

### IV. CONCLUSION

We have considered the first-order geometrical optics approximation for electromagnetic waves propagating in a smoothly inhomogeneous weakly anisotropic medium. Nontrivial competition occurs in such a medium between small polarization phenomena due to inhomogeneity and anisotropy. (In strongly anisotropic medium small polarization effects caused by the inhomogeneity can be neglected.) The quantum-mechanical formalism has enabled us to derive the equations of motion for translational and intrinsic degrees of freedom of the wave in a rigorous and consistent way.

The ray equations have been derived in the first approximation in small parameters of the geometrical optics and anisotropy. In contrast to the traditional zero-order approximation, the first-order corrections substantially depend on the waves polarization because of the spin-orbit interaction (optical Magnus effect or spin Hall effect of photons) and due to the medium anisotropy. While the optical Magnus effect depends only on the wave helicity, the corrections due to anisotropy contain, in general case, all components of the Stokes vector.

The equation of motion for the polarization or pseudospin is given both in Shrödinger- and Heisenberg-type representations. In the former one, it happens to be equivalent to the quasi-isotropic approximations equations [2], whereas in the latter representation it takes a simple form of the precession equation for the Stokes vector. The distinctive feature of anisotropic medium is that the polarization evolution is described by non-Abelian operator as it takes place in the evolution of electrons. It results in significant consequences. First, non-Abelian evolution leads to a lack of global basis of independent eigenmodes, i.e. to the energy exchange and mode conversion in any chosen basis. Second, owing to the interference of interacting modes the first-order corrections to the ray trajectory take the form of oscillations similar to *zitterbewegung* of electron with spin-orbit interaction.

The general theory has been illustrated by two systems with characteristic types of anisotropy. In gyrotropic magnetoactive medium the Faraday circular birefringence takes place, which is similar to circular birefringence due to the spin-orbit interaction of photons. As a result, the wave evolution remains Abelian just as in isotropic medium and the two effects are additive in the evolution of the polarization as well as in the deflections of the ray trajectories. It enables one to use magnetic field as an effective tool revealing or suppressing topological effects related to the spin-orbit interaction: the Rytov's polarization rotation (Berry phase) and optical Magnus effect. In a medium with linear birefringence of uniaxial crystal type (induced, e.g., by an external electric field) a non-Abelian polarization evolution and mode transformation take place. They arise from a simultaneous influence of the spin-orbit interaction (Berry phase) and medium anisotropy, and manifest themselves on any non-planar (e.g. helical) ray. At the wave propagation along such a ray the Stokes vector precesses about a certain direction, which can be regarded as periodic energy exchange between modes with different polarizations. This phenomenon can also cause oscillatory variations of the ray trajectory.

Finally, note that our approach allows generalization to the transverse waves in elastic media [10b] and to wave beams with optical vortices [27].

## ACKNOWLEDGEMENTS

The authors are grateful to M. Stelmaszczyk, Szczecin University of Technology, for generous help in preparing the text. This work was partially supported by Association Euratom-IPPLM, Poland (project P-12), STCU (grant P-307), and CRDF (grant UAM2-1672-KK-06).